# Integral refractive index imaging of flowing cell nuclei using quantitative phase microscopy combined with fluorescence microscopy

GILI DARDIKMAN,[1] YOAV N. NYGATE,[1] ITAY BARNEA,[1] NIR A. TURKO,[1] GYANENDRA SINGH,[1] BARHAM JAVIDI,[2] AND NATAN T. SHAKED[1,*]

[1] *Tel Aviv University, Faculty of Engineering, Department of Biomedical Engineering, Tel Aviv 69978, Israel*
[2] *University of Connecticut, Faculty of Engineering, Department of Electrical and Computer Engineering, Storrs 06269-4157, Connecticut, USA*
*\*nshaked@tau.ac.il*

**Abstract:** We suggest a new multimodal imaging technique for quantitatively measuring the integral (thickness-average) refractive index of the nuclei of live biological cells in suspension. For this aim, we combined quantitative phase microscopy with simultaneous 2-D fluorescence microscopy. We used 2-D fluorescence microscopy to localize the nucleus inside the quantitative phase map of the cell, as well as for measuring the nucleus radii. As verified offline by both 3-D confocal fluorescence microscopy and by 2-D fluorescence microscopy while rotating the cells during flow, the nucleus of cells in suspension that are not during division can be assumed to be an ellipsoid. The entire shape of a cell in suspension can be assumed to be a sphere. Then, the cell and nucleus 3-D shapes can be evaluated based on their in-plain radii available from the 2-D phase and fluorescent measurements, respectively. Finally, the nucleus integral refractive index profile is calculated. We demonstrate the new technique on cancer cells, obtaining nucleus refractive index values that are lower than those of the cytoplasm, coinciding with recent findings. We believe that the proposed technique has the potential to be used for flow cytometry, where full 3-D refractive index tomography is too slow to be implemented during flow.

**References and links**

1. C. M. Vest, *Holographic Interferometry* (Wiley, New York, 1979).
2. N. T. Shaked, "Quantitative phase microscopy of biological samples using a portable interferometer," Opt. Lett. **37**(11), 2016–2018 (2012).
3. N. T. Shaked, L. L. Satterwhite, M. T. Rinehart, and A. Wax, "Quantitative Analysis of Biological Cells Using Digital Holographic Microscopy," In *Holography, Research and Technologies*, J. Rosen Ed., (InTech, 2011), 219–236.
4. B. Rappaz, P. Marquet, E. Cuche, Y. Emery, C. Depeursinge, and P. J. Magistretti, "Measurement of the integral refractive index and dynamic cell morphometry of living cells with digital holographic microscopy," Opt. Express. **13**(23), 9361–9373 (2005).
5. B. Rappaz, F. Charrière, C. Depeursinge, P. J. Magistretti, and P. Marquet, "Simultaneous cell morphometry and refractive index measurement with dual-wavelength digital holographic microscopy and dye-enhanced dispersion of perfusion medium," Opt. Lett. **33**(7), 744–746 (2008).
6. B. Rappaz, A. Barbul, Y. Emery, R. Korenstein, C. Depeursinge, P. J. Magistretti, and P. Marquet, "Comparative study of human erythrocytes by digital holographic microscopy, confocal microscopy, and impedance volume analyzer, " Cytometry A **73**(10), 895–903 (2008).
7. C. L. Curl, C. J. Bellair, T. Harris, B. E Allman, P. J. Harris, A. G. Stewart, A. Roberts, K. A. Nugent, and L. Delbridge, "Refractive index measurement in viable cells using quantitative phase-amplitude microscopy and confocal microscopy," Cytometry A **65**(1), 88–92 (2005).

8. N. Cardenas, N. Ingle, L. Yu, and S. Mohanty, "Development of a digital holographic microscopy system integrated with atomic force microscope," Proc. SPIE 7904, 790409 (2011).
9. M. Balberg, M. Levi, K. Kalinowski, I. Barnea, S. K. Mirsky, and N. T. Shaked, "Localized measurements of physical parameters within human sperm cells obtained with wide-field interferometry," J. Biophotonics **10**(10), 1305–1314 (2017).
10. F. Charrière, A. Marian, F. Montfort, J. Kuehn, T. Colomb, E. Cuche, P. Marquet, and C. Depeursinge, "Cell refractive index tomography by digital holographic microscopy, " Opt. Lett. **31**(2), 178–180 (2006).
11. W. Choi, C. Fang-Yen, K. Badizadegan, S. Oh, N. Lue, R. R. Dasari, and M. S. Feld, "Tomographic phase microscopy," Nat. Methods **4**(9), 717–719 (2007).
12. M. Schürmann, G. Cojoc, S. Girardo, E. Ulbricht, J. Guck, and P. Muller, "Three-dimensional correlative single-cell imaging utilizing fluorescence and refractive index tomography," J. Biophotonics (2017).
13. F. Merola, P. Memmolo, L. Miccio, R. Savoia, M. Mugnano, A. Fontana, G. D'ippolito, A. Sardo, A. Iolascon, A. Gambale, and P. Ferraro, "Tomographic flow cytometry by digital holography," Light Sci. Appl. **6**(4), e16241 (2017).
14. M. M. Villone, P. Memmolo, F. Merola, M. Mugnano, L. Miccio, P. L. Maffettone, and P. Ferraro, "Full-angle tomographic phase microscopy of flowing quasi-spherical cells," Lab. Chip **18**(1), 126–131 (2018).
15. N. Lue, G. Popescu, T. Ikeda, R. R. Dasari, K. Badizadegan, and M. S. Feld, "Live cell refractometry using microfluidic devices," Opt. Lett. **31**(18), 2759–2761 (2006).
16. B. Kemper, S. Kosmeier, P. Langehanenberg, G. Von Bally, I. Bredebusch, and W. Domschke, J. Schnekenburger, "Integral refractive index determination of living suspension cells by multifocus digital holographic phase contrast microscopy," J. Biomed. Opt. **12**(5), 054009 (2007).
17. S. Kosmeier, B.Kemper, P. Langehanenberg, I. Bredebusch, J. Schnekenburger, A. Bauwens, and G. von Bally, "Determination of the integral refractive index of cells in suspension by digital holographic phase contrast microscopy," Proc. SPIE **6991**, 699110 (2008).
18. A. E. Ekpenyong, S. M. Man, S. Achouri, C. E. Bryant, J. Guck, and K. J. Chalut, "Bacterial infection of macrophages induces decrease in refractive index," J. Biophotonics **6**(5), 393–397 (2013).
19. M. Schürmann, J. Scholze, P. Müller, C. J. Chan, A. E. Ekpenyong, K. J. Chalut, and J. Guck, "Refractive index measurements of single, spherical cells using digital holographic microscopy," Methods Cell. Biol. **125**, 143–159 (2015).
20. Z. A. Steelman, W. J. Eldridge, J. B. Weintraub, and A. Wax, "Is the nuclear refractive index lower than cytoplasm? Validation of phase measurements and implications for light scattering technologies," J. Biophotonics **10**(12), 1714–1722 (2017).
21. M. Schürmann, J. Schloze, P. Müller, J. Guck, and C. J. Chan, "Cell nuclei have lower refractive index and mass density than cytoplasm," J. Biophotonics **9**(10), 1068–1076 (2016).
22. V. P. Tychinsky, A. V. Kretushev, I. V. Klemyashov, T. V. Vyshenskaya, N. A. Filippova, N. T. Raikhlin, and A. A. Shtil, "Quantitative real-time analysis of nucleolar stress by coherent phase microscopy," J. Biomed. Opt. **13**(6), 064032 (2008).
23. Y. Park, G. Popescu, K. Badizadegan, R. R. Dasari, and M. S. Feld,"Diffraction phase and fluorescence microscopy," Opt. Express **14**(18), 8263–8268 (2006).
24. E. Zlotek-Zlotkiewicz, S. Monnier, G. Cappello, M. Le Berre, and M. Piel, "Optical volume and mass measurements show that mammalian cells swell during mitosis," J. Cell Biol. **211**(4), 765–774 (2015).
25. N. Pavillon, A. Benke, D. Boss, C. Moratal, J. Kuhn, P. Jourdain, C. Depeursinge, P. J. Magistretti, and P. Marquet,"Cell morphology and intracellular ionic homeostasis explored with a multimodal approach combining epifluorescence and digital holographic microscopy," J. Biophotonics. **3**(7), 432–436 (2010).
26. M. Mir, Z. Wang, Z. Shen, M. Bednarz, R. Bashir, I. Golding, S. G. Prasanth, and G. Popescu, "Optical measurement of cycle-dependent cell growth," PNAS **108**(32), 13124–13129 (2011).
27. S. Chowdhury, W. J. Eldridge, A. Wax, and J. A. Izatt, "Structured illumination multimodal 3D-resolved quantitative phase and fluorescence sub-diffraction microscopy," Biomed. Opt. Express. **8**(5), 2496–2518 (2017).
28. R. A. Lotufo and E. R. Dougherty, *Hands-on Morphological Image Processing* (SPIE, Washington, 2003).
29. M. M. Villone, G. D'Avino, M. A. Hulsen, and P. L. Maffettone, "Dynamics of prolate spheroidal elastic particles in confined shear flow," Phys. Rev. E. **92**(6), 062303 (2015).
30. P. Girshovitz and N. T. Shaked, "Compact and portable low-coherence interferometer with off-axis geometry for quantitative phase microscopy and nanoscopy," Opt. Express **21**(5), 5701–5714 (2013).
31. P. Girshovitz and N. T. Shaked, "Real-time quantitative phase reconstruction in off-axis digital holography using multiplexing," Opt. Lett. **39**(8), 2262–2265 (2014).
32. D. C. Ghiglia and M. D. Pritt, *Two-Dimensional Phase Unwrapping: Theory, Algorithms, and Software* (Wiley, New York, 1998).

## 1. Introduction

Measurements of refractive index (RI) of biological cells *in vitro*, indicating their inner content and spatial arrangement, provide invaluable information for both medical diagnosis and biological research. Interferometric phase microcopy (IPM) enables imaging of isolated cells *in vitro*, without the use of cell staining, by measuring how much the light is delayed when passing through the sample rather than through a clear medium [1,2]. The phase delay measured not only provides good contrast without staining, but also consists of valuable information regarding both the internal geometrical structure and the RI distribution of the sample [3]. Nevertheless, the geometrical thickness and RI information are coupled in a way that makes it difficult to decipher each of these properties separately.

Several methods have been previously suggested for solving this RI-thickness coupling problem. These methods typically include performing two measurements, while changing either the illumination wavelength or the surrounding medium RI, creating two equations with two unknowns for each pixel, which enables the decoupling of the 2-D integral (thickness-average) RI profile from the thickness profile [4-6]. An alternative option is to measure the geometrical thickness directly (e.g., by atomic force microscopy (AFM) or confocal 3-D florescence microscopy) in addition to the IPM measurement, enabling the isolation of the integral RI of the cell [6-9]. Another option is using tomographic phase microscopy, which provides a means to not only decouple the information, but rather to obtain the 3-D distribution of the cell RI [10-12]. However, this method requires acquiring the sample from multiple perspectives, demanding that the angular scanning rate is faster than the sample dynamics, which might limit certain applications such as high-throughput imaging flow cytometry that requires analyzing thousands of cells per second. Although the random rolling of cells during flow can be exploited to obtain multiple perspectives [13,14], it demands tracking the cells and thus is limited by the full frame rate of the camera, while single exposure methods do not require cell tracking and thus allow faster flow rates for the same frame rate, mainly limited by the camera exposure time. In addition, tomography is prone to complications derived from the sample dynamics and the system stability, in comparison to single exposure methods.

Other approaches for solving the RI-thickness coupling problem are based on evaluating the local cell thickness under the reasonable assumption that cells in suspension assume a spherical shape [15-21]. A spherical model has also been used to evaluate the integral RI in the nucleus, both when it is isolated [20, 21], and in its native location inside the cell [22]. The latter method locates the nucleus position according to a local increase in phase thickness, a technique that may yield inaccurate results according to recent findings stating that the phase values in the area of the nucleus may be lower than that of the cytoplasm [11,12,20,21].

Several previous works suggested measuring fluorescence emission in concert with quantitative phase imaging in order to gain molecular specificity, and learn on biological processes such as cell growth and division [12, 21, 23-27]. It was also shown that IPM setups can be combined with fluorescence microscopy for obtaining simultaneous measurements [12, 23-27].

The present paper suggests integrating IPM with simultaneous 2-D fluorescence microscopy to localize the suspended cell nucleus in a coinciding quantitative phase image, and evaluate its radii. Then, we assume that both the entire suspended cell and its nucleus have shapes of ellipsoids, allowing their 3-D morphologies to be calculated based on the radii available from the 2-D fluorescent and IPM images. The cell and the nucleus 3-D morphologies can then be used to extract the integral RI of both the cytoplasm and the nucleus. To validate the assumption that the shape of the nucleus can be approximated as an ellipsoid, we use two independent experimental methods: confocal 3-D fluorescence microscopy and 2-D fluorescence microscopy while rotating the cell during flow.

Since the two online measurements, quantitative phase and 2-D fluorescence, are acquired simultaneously, the presented technique is useful for measurement the integral RI profile of cells during flow.

## 2. Methods

### 2.1 Theory

A transmission mode interferometric imaging system captures the phase difference between a beam that passes through a sample (a sample beam) and a beam that did not (a reference beam), by recording their interference pattern on a digital camera. The phase difference between the sample and the reference waves is proportional to the optical path difference (OPD) between them, as following:

$$\varphi(x,y) = \frac{2\pi}{\lambda} OPD(x,y), \tag{1}$$

where $\lambda$ is the illumination wavelength. Neglecting diffraction for simplicity, the OPD can be written as:

$$OPD(x,y) = \int_0^{h(x,y)} \left[ n(x,y,z) - n_m \right] dz, \tag{2}$$

where $z$ is the direction of light propagation, $h(x, y)$ is the thickness of the sample in the $z$ dimension, $n(x, y, z)$ is the RI distribution of the sample, and $n_m$ is the RI of the medium around the cell. In a discrete representation, Eq. (2) can be described as a finite sum:

$$OPD(i,j) = \sum_{k=1}^{N(i,j)} \left[ n(i,j,k) - n_m \right] \cdot \Delta z, \tag{3}$$

where $N(i, j)$ is the number of discrete increments defining the sample in the $z$ dimension for pixel $(i,j)$, and $\Delta z$ is the discrete increment length in the $z$ dimension, given by:

$$\Delta z = \frac{\Delta_{CCD}}{M}, \tag{4}$$

where $\Delta_{CCD}$ is the pixel size in the digital camera and $M$ is the total optical magnification used in the setup. Here, we assume that the $\Delta z$ increment size is the same as the $\Delta x$ and $\Delta y$ increment size.

Since $\Delta z$ and $n_m$ are constants and do not depend on $k$, we can take them out of the sum in Eq. (3). After multiplying and dividing by $N(i,j)$, we get:

$$OPD(i,j) = h(i,j) \cdot \left[ n_{cell}(i,j) - n_m \right], \tag{5}$$

where $h(i,j) = \Delta z \cdot N(i,j)$ is the thickness distribution of the cell and $n_{cell}(i,j) = [\sum_{k=1}^{N(i,j)} n(i,j,k)]/N(i,j)$ is the integral (thickness-average) RI distribution of the cell. Thus, the OPD in each pixel is the product of the thickness of the sample in that point and the difference between the integral RI of the sample in the $z$ dimension in that $(i,j)$ pixel and the medium, which constitutes the coupling problem.

In areas where the RI of the cell in the $z$ dimension is non-uniform, Eq. (5) can be further broken down into the contributions given by each organelle. For example, if the cell can be assumed to be comprised of a cytoplasm and a nucleus, we can write for the area of the OPD containing nucleus pixels (assuming that light propagating through the nucleus also goes through the cytoplasm) as follows:

$$OPD(i,j) = h_n(i,j) \cdot [n_n(i,j) - n_m] + h_c(i,j) \cdot [n_c(i,j) - n_m], \forall (i,j) \in \text{nucleus}, \quad (6)$$

where $h_n(i,j)$ and $n_n(i,j)$ are the thickness and integral RI distribution of the nucleus, respectively, and $h_c(i,j)$ and $n_c(i,j)$ are the thickness and integral RI distribution of the cytoplasm, respectively. Thus, we can express the average RI of the nucleus as follows:

$$n_n(i,j) = \frac{OPD(i,j) - h_c(i,j) \cdot [n_c(i,j) - n_m]}{h_n(i,j)} + n_m, \forall (i,j) \in \text{nucleus}, \quad (7)$$

where $OPD(i,j)$ and $n_m$ are known, but $h_c(i,j)$, $h_n(i,j)$ and $n_c(i,j)$ have to be evaluated.

Assuming that $h_n(i,j)$ and $h_c(i,j)$ are given or can be calculated, we can solve Eq. (7) by finding an estimate for the cytoplasm RI $n_c(i,j)$ from another area in the cell where there is only cytoplasm (outside the nucleus area), fulfilling:

$$n_c(i,j) = \frac{OPD(i,j)}{h_c(i,j)} + n_m, \forall (i,j) \in \text{cytoplasm}. \quad (8)$$

From the resulting RI profile, we can calculate the cytoplasm average RI value $n_{c,avg}$ by weighting the RI at each pixel with the corresponding height $h_c(i,j)$ [21] as follows:

$$n_{c,avg} = \frac{\sum_{i,j} n_c(i,j) \cdot h_c(i,j)}{\sum_{i,j} h_c(i,j)}, \forall (i,j) \in \text{cytoplasm}. \quad (9)$$

Assuming that the RI of the cytoplasm is largely homogenous, this value can then be used to estimate $n_c(i,j)$ in the area containing both nucleus and cytoplasm.

### 2.2 Thickness model

Evaluating the thickness profile of the nucleus $h_n(i,j)$ and the thickness profile of the entire cell $h_c(i,j)$ can be done by assuming ellipsoid models as shown in Fig. 1(a). A spherical model for an entire cell while not attached, in suspension or in flow, is a known paradigm [15-21], and the nucleus was also recently modeled as a sphere for thickness estimation [20-22]. Here, we assume that both the entire cell and the nucleus are ellipsoids rather than perfect spheres, practically allowing a more precise fit.

In order to build a thickness distribution for the nucleus and the cytoplasm, and also in order to know in which locations to apply Eq. (7) and Eq. (8), first the areas of the nucleus and the cytoplasm have to be located in the fluorescence and OPD images, respectively. To separate the cell area from the background in the OPD image and the nucleus area from the background in the fluorescent image, simple thresholds can be used, followed by morphological opening for noise removal [28]. After proper registration, the nucleus area can be detected on the OPD image. Once both areas are located, the minor and major axes radii ($R_1, R_2$) and the *x-y* orientation of the ellipse that has the same normalized second central moments as each area can be evaluated, as well as the centers of mass. Based on the minor and major axis lengths, the length of the radius in the *z* direction ($R_3$) can be calculated as the average for each area. A 3-D ellipsoid with the appropriate radii, *x-y* location and *x-y* orientation can then be computed for both the entire cell and the nucleus. In order to isolate the morphology of the cytoplasm area, the volume of the ellipsoid representing the nucleus can then be removed from that of the ellipsoid representing the entire cell. The thickness attributed to the nucleus or the cytoplasm at each pixel can then be calculated by integrating

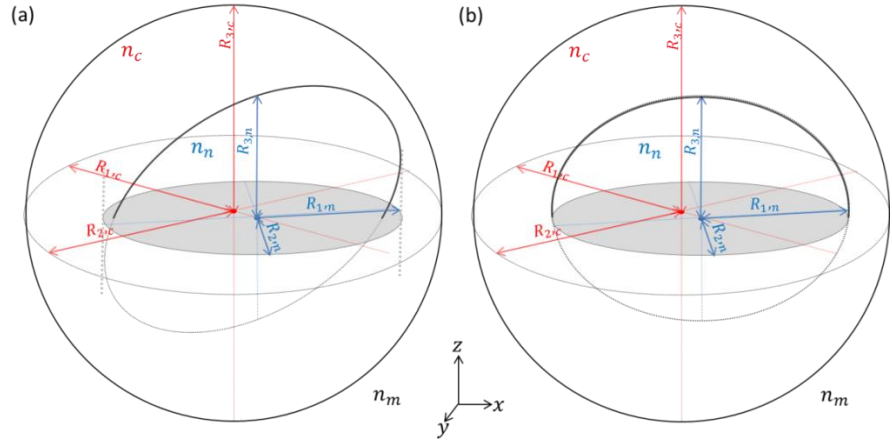

Fig. 1. A scheme of the cell and nucleus shape model: $n_n$, $n_c$ and $n_m$ are the RI of the nucleus, cytoplasm and medium, respectively. $R_{1,n}$ and $R_{2,n}$ are the radii in the *x-y* projection cast by the nucleus, and $R_{3,n}$ is the radius of the nucleus in the z dimension; $R_{1,c}$ and $R_{2,c}$ are the radii in the *x-y* projection cast by the entire cell, and $R_{3,c}$ is the radius of the entire cell in the *z* dimension. The radii in the *x-y* projection are extracted from the OPD image and fluorescence image for the cytoplasm and the nucleus, respectively, and the radius in the *z* dimension is estimated to be their average. (a) Nucleus modeled as a tilted ellipsoid. (b) Nucleus modeled as an ellipsoid with main axes aligned with Cartesian coordinates, with identical $R_{1,n}, R_{2,n}, R_{3,n}$. The red and blue circles denote the center of mass of the cytoplasm and nucleus, respectively. The Cartesian coordinate system indicates directions, where light propagates in the *z* direction.

$$h(i,j) = \Delta z \cdot \Sigma_k V(i,j,k), \qquad (10)$$

where $V(i,j,k)$ is the respective volume, $h(i,j)$ is the resultant thickness, and $\Delta z$ is given in Eq. (4).

As shown in Fig. 2, the thickness distribution calculated from Fig. 1(a) is nearly identical to the one calculated from Fig. 1(b). Thus, the actual orientation of the ellipsoid relative to the *z* axis is not significant for its thickness calculation.

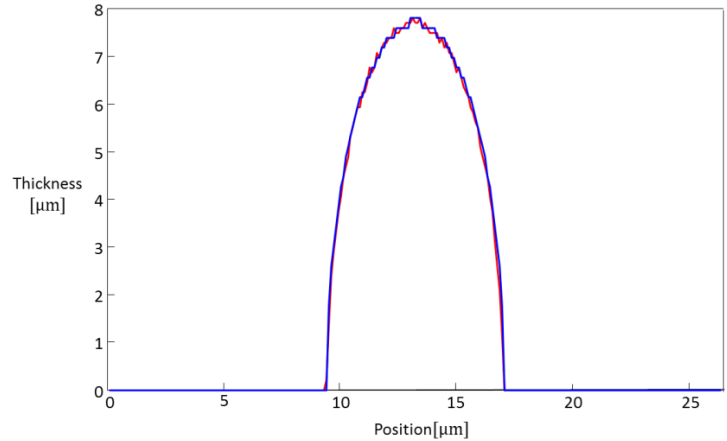

Fig. 2. Assessment of error due to assuming tilted versus untitled ellipsoid as the nucleus model. Red curve: cross section through the center of the thickness distribution of an ellipsoid tilted in 40° relative to the z axis, as shown schematically in Fig. 1(a). Blue curve: cross section through the center of the thickness distribution of an ellipsoid with axes aligned relative to the Cartesian coordinate system, as shown schematically in Fig. 1(b). Both ellipsoids have identical $R_{1,n}, R_{2,n}$ and $R_{3,n}$.

## 3. Results

### 3.1 Sample preparation

We imaged human colorectal adenocarcinoma colon cells, SW480 (CCL-228, ATCC). The cells were grown in DMEM medium supplemented with 10% fetal bovine serum, antibiotics and glutamine (Biological Industries, Beit HaEmeq, Israel). The cells were cultured at 37℃, 5% $CO_2$ in a humidified incubator. At the day of the experiment, the cells were harvested by trypsin solution and returned to another DMEM growth medium supplemented by Hoechst 33342 (Invitrogen, SN) staining solution at a concentration of 5 μg/ml for 30 minutes in 4℃.

### 3.2 Validation of nuclear morphology

To validate that the nucleus shape of a cell in suspension can be approximated as an ellipsoid, we first imaged cells with Hoechst-stained nuclei using a 3-D fluorescence confocal microscope (TCS SP5 II Confocal, Leica) with *z*-axis increments of 0.2 μm. For the confocal microscope measurement, unfixed cells were suspended in a well created by placing a sticker with a depth of 0.1 mm on top of a cover-slide. In order to prevent the cells from adhering to the cover slide, they were imaged less than 30 minutes after resting on the cover-slide, and placed in inhibiting conditions: at room temperature (rather than the optimal 37°C) and with a limited amount of air (rather than 5% CO2). No chemicals were used to avoid adhesion. As a result of letting the cells rest for several minutes after placing them in the imaging system, and due to the short scanning time (~2 minutes) and the low room temperature, the 3-D movement of the cells during the measurement is negligible. The result is given in Fig. 3.

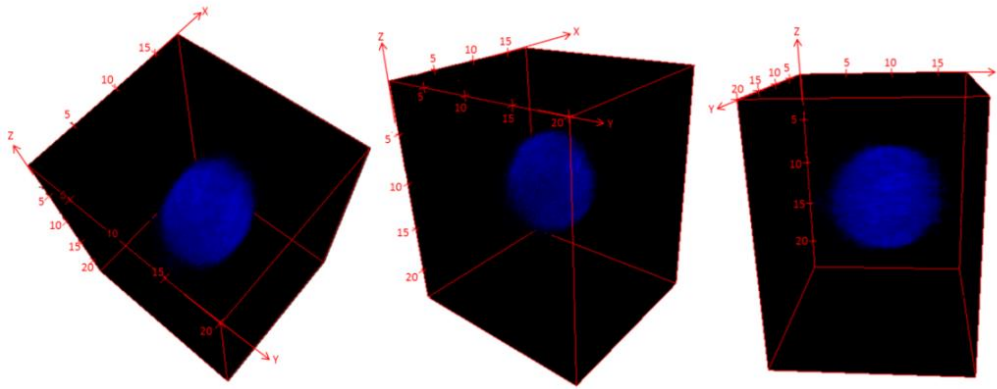

Fig. 3. 3-D rendering of a representative SW480 cancer cell nucleus, using 3-D fluorescence confocal microscopy, where the cell nucleus is labeled with Hoechst and the cell is not attached to the substrate. The numbers on the axes indicate length in μm.

To further demonstrate that SW480 nuclei maintain an approximate ellipsoid morphology during flow, we used a microfluidic channel (μ-slide VI0.1, IBIDI) and a syringe to induce cell rotation during flow, and tracked the morphology of 30 cells with stained nuclei under 2-D fluorescent microscopy (IX83 Inverted Microscope, Olympus). Due to very low confinement and a low flow rate during the fluorescent rotation experiment, the cells can be treated as rigid bodies [13,14,29]. A video of the rotation is given in Visualization 1, and an example of the nucleus of single cell during flow is given in Fig. 4. Note that as can be seen in Visualization 1, not all nuclei appear elliptic. This can be explained by the different phases of the cell lifecycle. During cell division, the nucleus morphology changes to allow the DNA to split between the two daughter cells. In any case, the 2-D fluorescence measurement can be used to select those cells with elliptic nuclei.

By observing the 2-D projections of an ellipsoid taken from a large enough number of random angles, the radii on its three main axes can be evaluated reliably. As a rule of thumb regarding the required number of projections during cell rotation, for a cell rotating at constant velocity, as few as 10 frames taken in equal temporal increments during a full revolution of the cell in a plane orthogonal to the camera would suffice to obtain an evaluation of the radii with an error of up to 5%. From each of the projections, the major and minor axis radii can be determined by fitting the area with an ellipse that has the same normalized second central moments. Then, the values measured for the minor and major axes radii from all projections of a single cell can be gathered into two vectors, respectively. The maximal value of the major axis radius vector evaluates the radius on the largest axis of the ellipsoid ($R_1$); the minimal value of the minor axis radius vector evaluates the radius of the smallest axis ($R_2$); and the minimal value of the major axis radius vector, which should be equal to the maximal value of the minor axis radius vector, evaluates the radius of the intermediate axis of the ellipsoid ($R_3$). Note that this equality relies on the assumption that the nucleus rotates around at least two orthogonal axes, as is usually the case due to the free rotation in the camera plane. Nevertheless, if the rotation is only around a single axis, one of the two vectors may have nearly constant values, and the evaluation should be made from the vector with the non-constant values. Even if none of the vectors are constant and the evaluations from both of them differ, taking the average value as the estimate still yields an error of less than 5%.

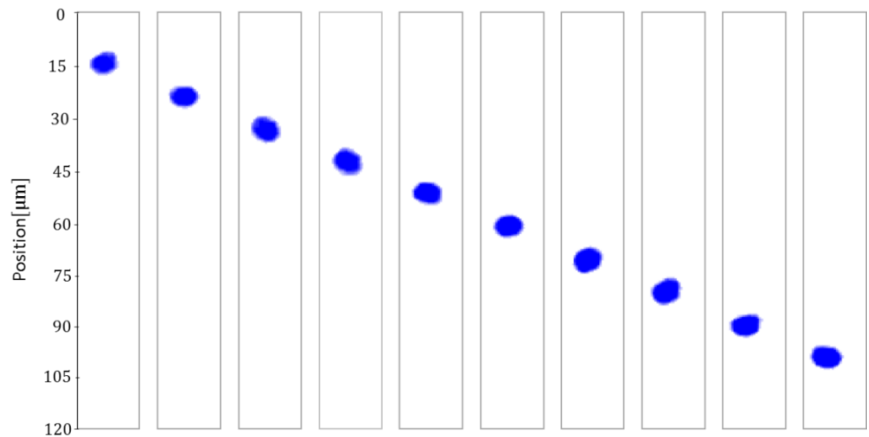

Fig. 4 2-D fluorescent tracking of the nucleus of a representative SW480 cell during flow.

We performed this analysis for 30 SW480 cell nuclei. At least 40 frames per cell were observed for each cell during rotation. The results of this analysis are summarized in Table 1. As can be seen from the standard deviation values, while the variability in the absolute sizes of the axes is relatively large, the ratios between the different axes exhibit a more similar proportion for all measured nuclei. Thus, in order to evaluate the predicted error induced by using the suggested model, we will use a normalized small axis with radius $R_2 = 1$ [AU], inducing $R_1 = 1.27$ [AU] and $R_3 = 1.13$ [AU] according to the average ratios, allowing the error to be calculated separately for each nucleus size by multiplying by the size of the smallest axis $R_2$.

**Table 1. Mean values for 30 SW480 nuclei. $R_1$, $R_2$ and $R_3$ are the main axes radii of the estimated ellipsoids.**

| $R_1 [\mu m]$ | $R_2 [\mu m]$ | $R_3 [\mu m]$ | $R_1/R_2$ | $R_1/R_3$ | $R_3/R_2$ |
|---|---|---|---|---|---|
| $4.55 \pm 0.57$ | $3.61 \pm 0.55$ | $4.05 \pm 0.56$ | $1.27 \pm 0.08$ | $1.13 \pm 0.06$ | $1.13 \pm 0.04$ |

In order to evaluate the predicted error in using the suggested model for the representative ellipsoid mentioned above, we must first know the sizes of the minor and the major axes in the projection, from which the third axis radius is calculated. If, for example, the nucleus happens to face the camera such that the largest radius $R_1 = 1.27$ [AU] and the smallest radius $R_2 = 1$ [AU] constitute the major and minor axes of the projection, our evaluation of the third axis radius as their average $R_3 = 1.135$ [AU] would be far more accurate than if $R_1$ and $R_3$ were to act as the main axes of the projection. In practice, the statistics of the orientation of the ellipsoid depends tightly on the flow regime, and is therefore hard to predict for the general case. To cover all cases, we assumed the two worst scenarios: either a combination of the largest axis with the intermediate axis, or a combination of the smallest axis with the intermediate axis. The results are given in Table 2 and Fig. 5.

**Table 2. Worst scenarios for using the ellipsoid model, calculated for normalized representative nucleus. All lengths are given normalized [AU]. The absolute values are given by multiplying by the size of the smallest axis.**

| Axis 1 radius | Axis 2 radius | Axis 3 radius evaluation | Axis 3 radius | Error |
|---|---|---|---|---|
| 1 | 1.13 | 1.065 | 1.27 | 0.205 |
| 1.27 | 1.13 | 1.2 | 1 | 0.2 |

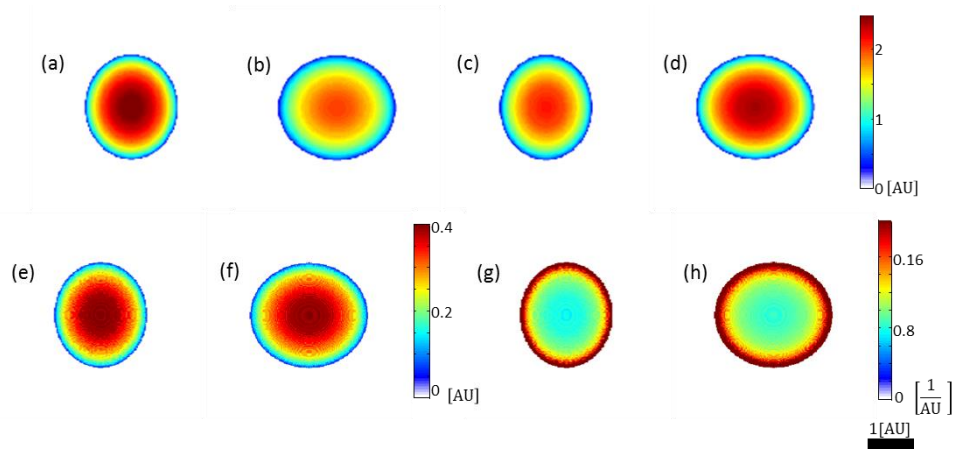

Fig. 5. Thickness distribution for worst orientation scenarios in using ellipsoid model, for a normalized representative cell nucleus. (a,c,e,g) Case 1, corresponding to the first row in Table 2. (b,d,f,h) Case 2, corresponding to the second row in Table 2. (a,b) True thickness distribution. (c,d) Evaluated thickness distribution. (e,f) Absolute value of error in thickness evaluation. (g,h) Absolute value of error in the inverse thickness evaluation.

The error in RI evaluation can then by evaluated using the following equation:

$$\Delta n(i,j) = n(i,j) - n_e(i,j) = \frac{OPD(i,j)}{h(i,j)} + n_m - \left[ \frac{OPD(i,j)}{h_e(i,j)} + n_m \right], \tag{11}$$

where $n(i,j)$ and $n_e(i,j)$ are the true and evaluated RI distributions, respectively, $h(i,j)$ and $h_e(i,j)$ are the true and evaluated height distributions, respectively, $n_m$ is the RI of the medium, and $OPD(i,j)$ is the measured OPD. This yields:

$$\Delta n(i,j) = OPD(i,j) \cdot \left[ \frac{1}{h(i,j)} - \frac{1}{h_e(i,j)} \right]. \tag{12}$$

The term multiplying the OPD profile in Eq. (12) can be seen in Figs. 5(g) and (h), such that the latter quantifies the error in RI up to a pixel-wise multiplication with the measured OPD and division by the size of the smallest axis. As can be seen from Figs. 5(g) and (h), the error is most significant in the edges, due to singularity in division by small numbers, thus locations closer to the center present higher reliability. For example, for an ellipsoid with a small axis radius size of 4 μm, the error in RI evaluation for a central pixel with an OPD value of 0.4 μm would be 0.0076 for the first case and 0.0086 for the second case, thus bounded by the latter for all possible orientations.

Note that by modeling the expected orientation of the ellipsoid for the specific flow regime used, one can modify the manner that the *z*-axis radius is estimated from the radii of the projection. For example, if the ellipsoid is expected to face the camera with its intermediate and major axes, one can evaluate the third radius as being equal to 1/1.13 = 0.88 of the minor axis of the projection, giving a better prediction than the average of the minor and major axes of the projection, thus improving the accuracy of the RI estimation.

*3.3 Combined IPM and fluorescence microscopy*

To demonstrate the proposed technique, we have combined an IPM system into a fluorescent microscope (IX83 Inverted Microscope, Olympus). In the proposed setup, shown in Fig. 6, a supercontinuum laser source (SuperK Extreme, NKT), coupled to an acousto-optic tunable filter AOTF (SuperK SELECT, NKT), emits red light (633±2.5nm). This light is reflected by mirror M1, and enters the inverted microscope. Inside the microscope, the beam passes through the sample, and is then magnified by microscope objective MO1 (Olympus UPlanFL N, 40×, 0.75 NA). The sample beam is then reflected by BS1 to exit the microscope, and projected by tube lens TL1 (f = 200 mm) on the image plane of the microscope, where a long-pass filter LP (cut off at 496 nm) is positioned. The sample beam is then magnified by a 4f lens configuration, built out of lenses L1 (f = 45 mm) and L2 (f =150 mm), and enters an external off-axis interferometric module [2, 30]. In this module, the magnified sample beam is split using beam splitter BS2. One of the beams is spatially filtered using lenses L3 (f = 150 mm) and L4 (f = 150 mm) and pinhole P (15 μm diameter), which effectively creates a reference beam that does not contain spatial modulation of the sample. The other beam from BS2 is projected through retro-reflector RR at a small angle, and together with the reference beam creates an off-axis hologram on Camera1 (Thorlabs, DCC1645C CMOS). At the same time, inside the microscope, white light (U-HGLGPS 130W Mercury burner, emission wavelength 360-770 nm) is filtered by an excitation filter Ex (350 nm with FWHM bandwidth of 50 nm), and reflected by dichroic mirror DM (long pass with cutoff at 400 nm). This beam is transmitted through BS1 and condensed by MO1 to illuminate the sample. The fluorescently labeled sample then emits blue light from the cell nucleus labeled with Hoechst 33342, which is magnified by MO1 and transmitted by BS1 and DM. This beam is projected through TL2 and filtered by the emission filter Em (460nm with FWHM bandwidth of 50 nm) to block red light, and projected onto Camera2 (Basler, acA2440-75um). This combined system allowed us to obtain interferograms and fluorescent images simultaneously without moving the sample, even if the sample is highly dynamic.

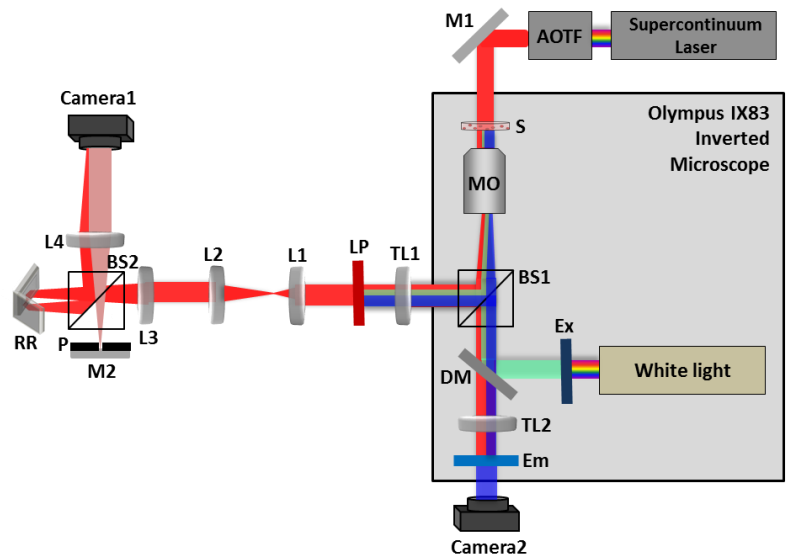

Fig. 6. Scheme of the combined IPM-epifluorescence setup used for imaging. IPM beams appear in red; The fluorescent excitation beam appears in green; The fluorescent emission beam appears in blue. Ex, excitation filter; DM, dichroic mirror; Em, emission filter; AOTF, acousto optical tunable filter; BS1, BS2, beam splitters; M1, M2, mirrors; RR, retroreflector; S, sample; MO1, microscope objective; TL1, TL2, L1, L2, L3, L4, lenses; LP, long-pass filter; Camera1, Camera2, digital cameras.

After the staining procedure (see Section 3.1) and prior to the imaging experiment, the suspended cells were inserted into a dish with a #1.5 cover-glass bottom (D35-20-1.5-N), and placed on the sample stage. In this demonstration, no fast flow or cell rotation was induced.

In order to perform image registration between the 2 modalities, fluorescent beads (6 μm micro particles, melamine resin, carboxylate-modified, FITC-marked, Fluka, Sigma-Adlrich) were used as a calibration target.

To extract the quantitative phase map from the acquired off-axis image interferograms, we used the Fourier transform filtering reconstruction algorithm [31], followed by the discrete cosine transform (DCT) based unweighted least squares (UWLS) algorithm for solving $2\pi$ phase ambiguities [32].

Figure 7 shows the different stages of applying the combined IPM-fluorescence processing method, as well as the results for the average RI in the nucleus area. First, Fig. 7(a) shows the reconstructed OPD image. The radii, calculated for the cell from fitting the cell boundary with an ellipse that has the same normalized second central moments, were 4.17 μm and 3.84 μm. Figure 7(b) shows the fluorescence image of the stained nucleus, and Fig. 7(c) shows the reconstructed OPD image with the nucleus boundary shown in broken line, found using the coinciding fluorescence image shown in Fig. 7(b). The radii, calculated for the nucleus by fitting the nucleus boundary with an ellipse that has the same normalized second central moments, were 3.27 μm and 2.99 μm. Figures 7(d) and 7(e) show the estimated thickness distribution for the cytoplasm and the nucleus, respectively, based on the morphology detected in Figs. 7(a) and 7(c) and the process explained in Section 2.2. Quantitatively, the relative mismatch of the elliptic fit is 4.5% of the pixels in the smallest rectangle containing the region for both the nucleus and the entire cell. Figure 7(f) shows the integral RI estimation for cytoplasm area, used to calculate a single value for the RI of the cytoplasm according to Eq. (9). Finally, Fig. 7(g) shows the resulting integral RI profile of the nucleus. Note that in Figs. 7(f) and 7(g), the RI estimation is calculated only for pixels that had non-zero thickness estimation, to avoid singularity.

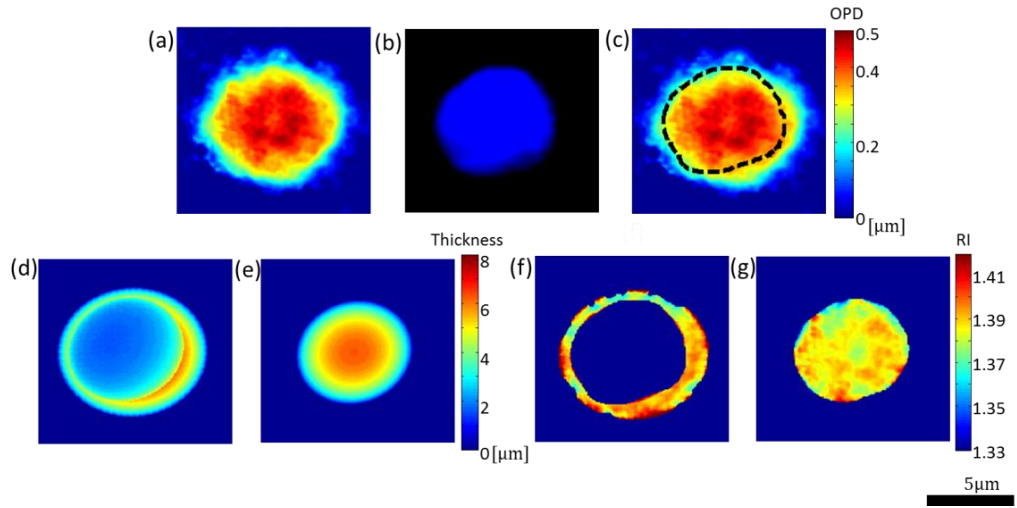

Fig. 7. Steps in extraction of the integral RI profile of a cell nucleus from the combined IPM/fluorescence measurement for a SW480 cell in suspension. (a) Reconstructed OPD image. (b) Fluorescence image of the labeled nucleus. (c) Reconstructed OPD image with nucleus boundary shown by broken line, as found from the fluorescence image. (d) Estimated thickness distribution for the cytoplasm. (e) Estimated thickness distribution for the nucleus. (f) Integral RI estimation for the cytoplasm only area. (g) The resulting integral RI profile of the cell nucleus.

Note that in contrast to tomography, that provides the 3-D RI profiles, the integral RI profiles shown in Figs. 7(f) and 7(g) are the average across the axial dimension. These integral RI profiles fit well with the recent observations that the RI inside the nucleus is lower than that in the cytoplasm, other than regions that may be attributed to the nucleoli [11,12,20,21].

## 4. Discussion and conclusions

We presented a new technique for measuring the nucleus integral RI distribution of suspended cells by combining IPM with simultaneous 2-D fluorescence microscopy, and demonstrated it on cancer cells. The results agree well with recent findings regarding the RI values of the nucleus. The suggested method can be applied on dynamics cells, such as cells during fast flow, where tomographic phase microscopy for 3-D RI acquisition is not fast enough. Our technique is based on single-exposure imaging in two parallel cameras, and thus the acquisition rate can be equal to the camera framerate (up to several thousands of frames per second) times the number of cells in the field of view. The only limitations, which are general demands for imaging flow cytometry, are that the cells can be approximated as rigid bodies, and avoiding motion blur. For the latter, if using a camera with an exposure time of 1 µsec, the cells can traverse up to 1 pixel each µsec.

Although the extracted parameter is the integral (thickness-average) RI profile, and it may blur out certain details such as the nucleoli in comparison to the 3-D RI profile that can be retrieved from tomographic phase microscopy measurement, it might be the only option for cells flowing faster than the tomographic phase microscopy acquisition rates. In any case, it is still superior to having the coupled thickness-RI phase profile.

Even though this method uses ellipsoid models to fit a broad range of morphologies, it is most accurate when both the cell and the nucleus are perfect spheres, as in such case the evaluation of the third radius as the average of those restored from the *x-y* projection is always accurate. While cells in suspension have been previously proven to be nearly spherical, this is not the case for their nuclei. To check the validity of the ellipsoid model for the nucleus and predict the error in evaluation of its third radius, we acquired the 3-D morphology of the nucleus for SW480 cancer cells by confocal 3-D fluorescence microscopy. Furthermore, we have analyzed 2-D fluorescence microscopy images of flowing cells during rotation, and discovered that many nuclei of suspended cells are indeed ellipsoids, yielding a representative ellipsoid structure with characteristic aspect ratio. We then calculated the predicted error considering the worst case in terms of the ellipsoid orientation relative to the camera plane during flow. Note that by modeling the expected orientation of the ellipsoid for the specific flow regime used, one can modify the manner that the $z$ dimension radius is estimated from the radii of the projection, thus reducing the maximal predicted error.

While the model suggested in this manuscript can be easily applied to many types of cells, we suggest performing similar preliminary analysis to verify the ellipsoidal shape of the nucleus and cell, in order to evaluate the predicated maximal error for the specific cell type used. This prior analysis has to be performed once per cell type and thus it does not affect the capability of the method to achieve high temporal resolution in the actual measurement.

Overall, we believe that the method presented here has the potential of having great clinical importance, providing the cell nucleus RI profile during imaging flow cytometry.

**Funding**

Horizon 2020 European Research Council (ERC) (678316); U.S–Israel Binational Science Foundation (BSF) (2013341); Tel Aviv University Center for Light-Matter Interaction.

**Disclosures**

The authors declare that there are no conflicts of interest related to this article.